\documentclass[fleqn,12pt]{article}
\usepackage{msrtr}\msrtrno{2003-08}

\def \OpusText       {KRML 118 / VK0099 / MSR-TR-2003-08}
\def \DocumentTitle  {On computing the fixpoint of a set of boolean equations}
\def \RevisionDate   {30 December 2003}
\newcommand{\comment}[1]{}

\usepackage{times}
\usepackage{Rustan}
\usepackage{amssymb}
\usepackage{daytime} % vkuncak

\begin{document}

\bibliographystyle{plainsort}

%%%%%%%%%%%%%%%%%%%%%%%  MATH CODES  %%%%%%%%%%%%%%%%%%%%%%%%%%%%%%%%%%

\mathsurround=2pt

\newenvironment{mmath}{\[\begin{array}{@{}l@{}}}{\end{array}\]}
\newenvironment{labelmath}[1]%
        {\begin{equation}\label{#1}\begin{array}{@{}l@{}}}%
        {\end{array}\end{equation}}
\makeatletter
\newcommand{\mmathlabel}[1]{%
                                \refstepcounter{equation}%
                                \label{#1}%
                                \rlap{\kern-\@totalleftmargin\kern-2.5em%
                                     \hbox to\hsize{\hfil(\arabic{equation})}}%
                                \ignorespaces
                             }
\makeatother

%%%%%%%%%%%%%%%%%%%%%%%  MATH SYMBOLS  %%%%%%%%%%%%%%%%%%%%%%%%%%%%%%%%

\newcommand{\atmost}{\leq}
\newcommand{\atleast}{\geq}

\newcommand{\MathOp}[2]{{}\mathbin{\hbox{$\mkern#2mu#1\mkern#2mu$}}{}}

\newcommand{\Equiv}{\MathOp{\equiv}{6}}
\newcommand{\Equal}{\MathOp{=}{6}}
\newcommand{\Xor}{\MathOp{\not\equiv}{6}}
\newcommand{\imp}{\Rightarrow}
\newcommand{\expl}{\Leftarrow}
\newcommand{\Imp}{\MathOp{\imp}{4}}
\newcommand{\Exp}{\MathOp{\expl}{4}}
\newcommand{\And}{\MathOp{\wedge}{2}}
\newcommand{\Or}{\MathOp{\vee}{2}}
\newcommand{\Neg}{\neg}

%%%%%%%%%%%%%%%%%%%%%%%  THEOREMS AND THE LIKE  %%%%%%%%%%%%%%%%%%%%%%%

\newcommand{\EndProofBullet}{\makebox[5pt][l]{
}}

%%%%%%%%%%%%%%%%%%%%%%%  CALCULATION RELATED  %%%%%%%%%%%%%%%%%%%%%%%%%

\newcommand{\CalcD}[1]{\begin{description}
                \item \begin{tabbing}\qquad\=\qquad\=\kill
                \>#1\end{tabbing}
                \end{description}}
\newcommand{\Calc}[1]{\begin{tabbing}\qquad\=\qquad\=\kill
                \>#1\end{tabbing}}
\newcommand{\nlcalc}{$\\*\>$}
\newcommand{\conn}[2]{\\*$#1$\>\>%
    \def\hint{#2}
    \ifx\hint\empty
    \else
        \{\phantom{i}\hint\phantom{i}\}
    \fi
    \\\>}
\newcommand{\Conn}[1]{\conn{=}{#1}}
\newcommand{\nlhint}{\\*\>\>\phantom{\{i}\ \ }

% The rest of this section due to Marcel van der Goot

\makeatletter
\newcommand{\calcEqLabel}[1]{%
                                \refstepcounter{equation}%
                                \label{#1}%
% IF you want equation numbers at the left-hand side:
%                               \llap{\rlap{(\arabic{equation})}%
%                                     \kern\@totalleftmargin\qquad}%
% ELIF you want equation numbers at the right-hand side:
                                \rlap{\kern-\@totalleftmargin\kern-2em%
                                     \hbox to\hsize{\hfil(\arabic{equation})}}%
% FI
                                \ignorespaces
                             }
\newcommand{\calcComment}[1]{%
                                \rlap{\kern-\@totalleftmargin\kern-10em%
                                     \hbox to\hsize{\hfil #1}}%
                                \ignorespaces
                             }
\makeatother

%%%%%%%%%%%%%%%%%%%%%%%  QUANTIFIERS  %%%%%%%%%%%%%%%%%%%%%%%%%%%%%%%%%

% The following takes care of the spacing before the close quantification
% bracket.
% Local use only
\newcommand{\quantOpen}[1]{#1\,}
% The following \testquant hack due to Marcel van der Goot.
\newcommand{\quantClose}[1]{\mskip\thickmuskip#1\futurelet\next\testquant}
\newcommand\testquant{
        \ifx\next\quantClose \mskip-\thickmuskip \fi
        \ifx\next\QuantClose \mskip-\thickmuskip \fi}
% End Local use only

\newcommand{\DummyRangeSepSymbol}{\left\bracevert\phantom{i}\right.\!\!}
\newcommand{\RangeTermSepSymbol}{\bullet}

\newcommand{\DummyRangeSep}{\DummyRangeSepSymbol}
\newcommand{\RangeTermSep}{\RangeTermSepSymbol~}
\newcommand{\DummyTermSep}{\RangeTermSepSymbol~}

\newcommand{\QuantOpen}[1]{\quantOpen{(}#1\,}
\newcommand{\QuantClose}{\quantClose{)}}

\newcommand{\QuantExpr}[4]{\QuantOpen{#1}
                #2
                \def\range{#3}
                \ifx\range\empty
                  \DummyTermSep
                \else
                  \DummyRangeSep#3\RangeTermSep
                \fi
                #4\QuantClose}

\newcommand{\Set}[3]{\quantOpen{\{}
                #1
                \def\range{#2}
                \ifx\range\empty
                  \DummyTermSep
                \else
                  \DummyRangeSep#2\RangeTermSep
                \fi
                #3\quantClose{\}}}

%%%%%  Particular quantifiers

\newcommand{\qForall}{\forall}
\newcommand{\Forall}[3]{\QuantExpr{\qForall}{#1}{#2}{#3}}

\newcommand{\qExists}{\exists}
\newcommand{\Exists}[3]{\QuantExpr{\qExists}{#1}{#2}{#3}}

%%%%%%%%%%%%%%%%%%%%%%%  SECTION RELATED  %%%%%%%%%%%%%%%%%%%%%%%%%%%%%

\newcommand{\Section}[1]{\section{#1} \setcounter{subsection}{-1}}
\newcommand{\SubSection}[1]{\subsection{#1} \setcounter{subsubsection}{-1}}

\setcounter{section}{-1}
\setcounter{table}{-1}
\setcounter{figure}{-1}
\setcounter{equation}{-1}

\newenvironment{Enumerate}
                {\begin{enumerate} \setcounter{enumi}{-1}}
                {\end{enumerate}}

\renewcommand{\thefootnote}{\fnsymbol{footnote}}

%%%%%%%%%%%%%%%%%%%%%%%  LATIN ABBREVIATIONS  %%%%%%%%%%%%%%%%%%%%%%%%%

\def\eg{{\it e.g.}}
\def\ie{{\it i.e.}}
\def\viz{{\it viz.}}
\def\etc{{\it etc.}}
\def\vs{{\it vs.}}
\def\cf{{\it cf.}}
\def\etal{{\it et al.}}
\def\homedir{\~{ }}

% Lyle's unbreakable hyphen
\newcommand{\unbreakableHyphen}{\setbox0=\hbox{-}\setbox1=\hbox{-\/}%
        \kern\wd0\kern-\wd1\hbox{-}}

% end of krml.tex

%  macros used in this paper
\newcommand{\Bool}{\mathbb{B}}
\newcommand{\DomA}{\mathbb{A}}
\newcommand{\tup}[1]{\overrightarrow{#1}}
\newcommand{\tupatmost}{\mathrel{\tup{\atmost}}}
\newcommand{\comp}{\circ}
\newcommand{\LambdaExpr}[3]{\QuantExpr{\lambda}{#1}{#2}{#3}}
\newcommand{\zerotup}{\tup{0}}
\newcommand{\zerofunc}{\dot{0}}
\newtheorem{theorem}{Theorem}
\newtheorem{lemma}[theorem]{Lemma}
\newtheorem{corollary}[theorem]{Corollary}
\setcounter{theorem}{-1}
\newcommand{\Gets}{\,{:=}\;}
\newcommand{\Case}[1]{\par\noindent \textsc{Case $#1$:~~}}
\newcommand{\SubCase}[1]{\par\noindent \textsc{Sub-case $#1$:~~}}
\newcommand{\Proof}{\par\noindent\textit{Proof.}}
\newcommand{\ProofOf}[1]{\par\noindent\textit{Proof of #1.}}
\newcommand{\EndProof}{\qquad\EndProofBullet}

\newcommand{\foo}{\mathfrak{f}}
\newcommand{\goo}{\mathfrak{g}}
\newcommand{\hoo}{\mathfrak{h}}

\newcommand{\down}[2]{\QuantExpr{{\downarrow}}{#1}{}{#2}}
% Note, in order to write \'c in the tabbing environment, one
% needs to write it as {\accent"13 c}.
\newcommand{\BL}{Beki{\accent"13 c}-Leszczy{\l}owski}
\newcommand{\Let}{\mathbf{let}~}
\newcommand{\In}{~\mathbf{in}~}

%  end of macros used in this paper

\setcounter{page}0

\title{\DocumentTitle}
\author{Viktor Kuncak\\\normalsize MIT\\\normalsize\texttt{vkuncak@mit.edu} \and
        K. Rustan M. Leino\\\normalsize Microsoft Research\\\normalsize\texttt{leino@microsoft.com}}
\date{\RevisionDate}

\begin{abstract}
  This paper presents a method for computing a least fixpoint of a
  system of equations over booleans.  The resulting computation can
  be significantly shorter than the result of iteratively evaluating
  the entire system until a fixpoint is reached.
\end{abstract}

\maketitle

\Section{Introduction}

Many problems in computer science, in particular those arising in the
context of program analysis, involve the computation of a least (or,
dually, greatest) fixpoint of a system of equations.  In this paper,
we consider a way to compute a least fixpoint when the equations
involved are over the booleans.  In some important cases, the
resulting computation can be significantly shorter than the
computation that iteratively evaluates the entire system until a
fixpoint is reached.

Let us begin with an overview of our result.  We restrict our
attention to a finite lattice.  A finite lattice is a complete
lattice and has no infinite ascending chains, and any monotonic
function on such a lattice is also continuous.  Hence, the Kleene
Fixpoint Theorem~\cite{Kleene:fixpoint} states that the least fixpoint
of any monotonic function $F$ is the lattice join of the sequence of
elements
\[  F^0(\bot),~ F^1(\bot),~ F^2(\bot),~ \ldots  \]
where exponentiation denotes successive function applications and
$\bot$ denotes the bottom element of the lattice.  Because this
sequence is ascending and because the lattice is finite, % we have that 
there exists a natural number $K$ such that
\[  F^K(\bot)  \]
is the least fixpoint of $F$.  We call the least such $K$ the
\emph{fixpoint depth} of $F$.

If we are able to evaluate function $F$ and if we are able
to determine whether two given lattice elements are equal,
then we can compute the least fixpoint of $F$: starting from
the value $\bot$, repeatedly apply $F$ until the application
of $F$ leaves the value unchanged.  The existence of a
fixpoint depth guarantees that this process terminates.  In
this paper, we consider the problem of computing an
\emph{expression} for the least fixpoint, without computing
the \emph{value} of the expression.  By first computing a small
expression for the least fixpoint, we can relegate the
computation of the value of the expression to an external
tool such as a SAT solver \cite{Leino:SPIN2003}.  In the
sequel we therefore do not assume that we are able to
compute the value of an expression into a particular
lattice element.

The fixpoint depth of a function $F$ on a lattice is bounded by the
height of the lattice.  Therefore, for the 2-element lattice $\Bool$
of the booleans (which has height 1), the least fixpoint of $F$ is
given by $F(\bot)$, and for the $2^n$-element lattice $\Bool^n$ that is
the Cartesian product space of $n$ booleans (which has height $n$),
the least fixpoint of $F$ is given by $F^n(\bot)$.

Any function $F \colon \Bool^n \to \Bool^n$ can be represented
isomorphically by $n$ functions $f_i \colon \Bool^n \to \Bool$.  We
write
\[  F  \Equal  (f_1,\ldots,f_n)  \]
where the tuple of functions is itself defined to be a function, as
follows, for any $n$-tuple $X$ of booleans:
\[
  (f_1,\ldots,f_n)(X)
  \Equal
  (f_1(X),~ \ldots,~ f_n(X))
\]
For example, let $n=3$ and let $F=(f,g,h)$.  Then, the least fixpoint
of $F$ equals $F^3(\bot,\bot,\bot)$, as we have argued above.  In
terms of the functions $f,g,h$, this expands to:
\[\begin{array}{@{}l@{}l@{}l@{}}
( & f( & f( f(\bot,\bot,\bot),~ g(\bot,\bot,\bot),~ h(\bot,\bot,\bot)), \\
  &    & g( f(\bot,\bot,\bot),~ g(\bot,\bot,\bot),~ h(\bot,\bot,\bot)), \\
  &    & h( f(\bot,\bot,\bot),~ g(\bot,\bot,\bot),~ h(\bot,\bot,\bot))), \\
  & g( & f( f(\bot,\bot,\bot),~ g(\bot,\bot,\bot),~ h(\bot,\bot,\bot)), \\
  &    & g( f(\bot,\bot,\bot),~ g(\bot,\bot,\bot),~ h(\bot,\bot,\bot)), \\
  &    & h( f(\bot,\bot,\bot),~ g(\bot,\bot,\bot),~ h(\bot,\bot,\bot))), \\
  & h( & f( f(\bot,\bot,\bot),~ g(\bot,\bot,\bot),~ h(\bot,\bot,\bot)), \\
  &    & g( f(\bot,\bot,\bot),~ g(\bot,\bot,\bot),~ h(\bot,\bot,\bot)), \\
  &    & h( f(\bot,\bot,\bot),~ g(\bot,\bot,\bot),~ h(\bot,\bot,\bot))))
\end{array}\]
We refer to this closed form of the fixpoint as the \emph{Expanded
  Closed Form}.
A different way to write down the Expanded Closed Form, which shares
  common subexpressions, is:
\[
\begin{array}{@{}l@{}l@{~}l@{}l@{~}l@{}l@{~~}l@{}}
  \Let a_1 & = \bot, &
       a_2 & = \bot, &
       a_3 & = \bot & \In \\
  \Let b_1 & = f(a_1,a_2,a_3), &
       b_2 & = g(a_1,a_2,a_3), &
       b_3 & = h(a_1,a_2,a_3) & \In \\
  \Let c_1 & = f(b_1,b_2,b_3), &
       c_2 & = g(b_1,b_2,b_3), &
       c_3 & = h(b_1,b_2,b_3) & \In \\
  \Let d_1 & = f(c_1,c_2,c_3), &
       d_2 & = g(c_1,c_2,c_3), &
       d_3 & = h(c_1,c_2,c_3) & \In \\
  \multicolumn{2}{l}{
  \quad (d_1,d_2,d_3)}
\end{array}
\]
This representation is cubic in $n$, which means that computing it
may take time and space that is cubic in $n$.\footnote{%
  If we allow ourselves to write functions of $n$ arguments as
  functions over $n$-tuples, then we can obtain a quadratic
  representation.  For example, with $n=3$, we have
  $
  \Let a_1=\ldots, a_2=\ldots, a_3=\ldots \In \Let a=(a_1,a_2,a_3) \In
  \Let b_1=f(a), b_2=g(a), b_3=h(a) \In \Let b=(b_1,b_2,b_3) \In
  \ldots
  $.
}

Let us consider another closed form, which we call the \emph{Pruned
  Closed Form}.  In the Pruned Closed Form, an application of a
function $f_i$ is replaced by $\bot$ if it occurs in another
application of the same function $f_i$.  For
the example above, where $n=3$, the Pruned Closed Form is:
\[
\begin{array}{@{}l@{}l@{}l@{}l@{}l@{}l@{}}
( & f( & \bot, \\
  &    & g( \bot,~              & \bot,~              & h(\bot,\bot,\bot)&), \\
  &    & h( \bot,~              & g(\bot,\bot,\bot),~ & \bot&)), \\
  & g( & f( \bot,~              & \bot,~              & h(\bot,\bot,\bot)&), \\
  &    & \bot, \\
  &    & h( \bot,~              & \bot,~              & \bot&)), \\
  & h( & f( \bot,~              & g(\bot,\bot,\bot),~ & \bot&), \\
  &    & g( \bot,~              & \bot,~              & \bot&), \\
  &    & \bot &&&\;))
\end{array}
\]

If we do not have any interpretation for the functions $f_i$---in other
words, if each $f_i$ is just a symbolic name for an uninterpreted
function---then the cubic-sized Expanded Closed Form may be a
reasonably small closed-form representation of the fixpoint.  The
Pruned Closed Form is generally much larger than cubic in $n$: for
every subset $S$ of $f_2,\ldots,f_n$, function $f_1$ appears expanded
in a context where the set of enclosing functions is $S$.  (A smaller
Pruned Closed Form can be obtained by taking advantage of common
subexpressions.)  However, there are cases where the Pruned Closed
Form can be significantly smaller than the Expanded Closed Form, for
example when the fixpoint computation is dominated by the computation
of local fixpoints, meaning fixpoints that involve only a small number
of the functions.  An important situation in program analysis where
this case applies is when each function represents a control point in
a given program, a function is defined in terms of the functions
corresponding to the successor (or predecessor) control points, and
the given program contains many local loops.

For example, suppose
\[
  f(x,y,z) \Equal \foo(x,y)
\qquad
  g(x,y,z) \Equal \goo(x)
\qquad
  h(x,y,z) \Equal \hoo(y,z)
\]
for some functions $\foo$, $\goo$, and $\hoo$.  Then the Expanded
Closed Form is
%\[
% (\begin{array}[t]{@{}l@{}l}
%    \foo( & \foo(\foo(\bot,\bot),\goo(\bot)), \\
%          & \goo(\foo(\bot,\bot)), \\
%          & \hoo(\goo(\bot),\hoo(\bot,\bot))), \\
%    \goo( & \foo(\foo(\bot,\bot), \goo(\bot))), \\
%    \hoo( & \goo(\foo(\bot,\bot)), \\
%          & \hoo(\goo(\bot),\hoo(\bot,\bot))))
%        \end{array}
%\]
\[
\begin{array}{@{}l@{}l@{~}l@{}l@{~}l@{}l@{~~}l@{}}
  \Let a_1 & = \bot, &
       a_2 & = \bot, &
       a_3 & = \bot & \In \\
  \Let b_1 & = \foo(a_1,a_2), &
       b_2 & = \goo(a_1), &
       b_3 & = \hoo(a_2,a_3) & \In \\
  \Let c_1 & = \foo(b_1,b_2), &
       c_2 & = \goo(b_1), &
       c_3 & = \hoo(b_2,b_3) & \In \\
  \Let d_1 & = \foo(c_1,c_2), &
       d_2 & = \goo(c_1), &
       d_3 & = \hoo(c_2,c_3) & \In \\
  \multicolumn{2}{l}{
  \quad (d_1,d_2,d_3)}
\end{array}
\]
In contrast, the Pruned Closed Form yields the much shorter expression
\[
 (\begin{array}[t]{@{}l@{}l}
    \foo( & \foo(\bot,\goo(\bot)), \\
          & \goo(\bot)), \\
    \goo( & \foo(\bot, \bot)), \\
    \hoo( & \goo(\foo(\bot,\bot)), \\
          & \bot))
        \end{array}
\]
More generally, for an even $n$, suppose $f_i(x_1,\ldots,x_n)$ is
$\foo_i(x_i, x_{i+1})$ when $i$ is odd and $\foo_i(x_{i-1}, x_i)$ when
$i$ is even.  Then the Expanded Closed Form is still cubic, whereas the Pruned
Closed Form is the linear-sized expression
\[\begin{array}{ll}
  (
  &      \foo_1(\bot, \foo_2(\bot, \bot)),
  \quad  \foo_2(\foo_1(\bot, \bot), \bot),
  \\ &   \ldots
  \\ &   \foo_i(\bot, \foo_{i+1}(\bot, \bot)),
  \quad  \foo_{i+1}(\foo_i(\bot, \bot), \bot),
  \\ &   \ldots
  \\ &   \foo_{n-1}(\bot, \foo_n(\bot, \bot)),
  \quad  \foo_n(\foo_{n-1}(\bot, \bot), \bot)
  \\ )
\end{array}\]

In the rest of this paper, we define the Pruned Closed Form more
precisely and prove that it yields the same value as the Expanded
Closed Form.

\Section{Using the \BL{} Theorem}  \label{sec:bl}

In this section, we sketch how to obtain the Pruned Closed Form by
applications of the \BL{}
Theorem~\cite{Bekic:theorem,Leszczylowski:theorem}.

We write
\[  \down{x}{R(x)}  \]
for the lattice meet of all values for $x$ that satisfy the predicate
$R(x)$.  For any monotonic function $F$, we then write
\begin{equation}\label{fixpoint:notation}
  \down{x}{x = F(x)}
\end{equation}
to denote the least fixpoint of $F$, because the Tarski Fixpoint
Theorem~\cite{Tarski:theorem} says that the meet of all fixpoints is
itself a fixpoint.
Using for a function $F \colon \Bool^n \to \Bool^n$ the isomorphic
representation of $n$ functions $f_i \colon \Bool^n \to \Bool$, we
can write (\ref{fixpoint:notation}) equivalently as:
\[\begin{array}{@{}ll@{}c@{}l@{}l@{}}
  \down{x_1,\ldots,x_n}{
  & x_1 & \Equal & f_1(x_1,\ldots,x_n) & \And \\
  &     & \vdots &                & \And \\
  & x_n & \Equal & f_n(x_1,\ldots,x_n)}
\end{array}\]

We can now state the \BL{}
Theorem~\cite{Bekic:theorem,Leszczylowski:theorem}, for any monotonic
functions $F$ and $G$ (possibly over different lattices):
\[\begin{array}{@{}ll@{}}
  & \down{a,b}{a = F(a,b) \And b = G(a,b)} \\
  = \\
  & \down{a,b}{a = F(a,b) \And b = \down{b}{b=G(a,b)}}
\end{array}\]
Note that each side of the equality expresses a fixpoint in the
lattice $\Bool^n$ if $F$ and $G$ are functions of types $\Bool^p \times
\Bool^q \to \Bool^p$ and $\Bool^p \times
\Bool^q \to \Bool^q$, respectively, for $p$ and $q$ such that $p+q=n$.

A consequence of the \BL{} Theorem and the Kleene Fixpoint Theorem for a
known fixpoint depth is the following lemma:
\begin{lemma}\label{BLK:lemma}
For any lattice domain $\DomA$ and monotonic functions
$F \colon \DomA \times \Bool \to \DomA$ and
$G \colon \DomA \times \Bool \to \Bool$,
\[\begin{array}{@{}ll@{}}
  & \down{a,b}{a = F(a,b) \And b = G(a,b)} \\
  = \\
  & \down{a,b}{a = F(a,b) \And b = G(a,\bot)}
\end{array}\]
\end{lemma}
\Proof{}
\Calc{
  $ \down{a,b}{a = F(a,b) \And b = G(a,b)} $
\Conn{ \BL{} Theorem }
  $ \down{a,b}{a = F(a,b) \And b = \down{b}{b=G(a,b)}} $
\Conn{ $\LambdaExpr{b}{}{G(a,b)}$ is a function on $\Bool$, and
       \nlhint therefore its fixpoint depth is at most 1, and
       \nlhint therefore $\down{b}{b=G(a,b)} \Equal G(a,\bot)$ }
  $ \down{a,b}{a = F(a,b) \And b = G(a,\bot)} $
\EndProof
}

Using Lemma~\ref{BLK:lemma}, we now show that the Pruned Closed Form
is indeed the least fixpoint in $\Bool^2$.  For any monotonic boolean
functions $f$ and $g$:
\Calc{
\calcEqLabel{spec:thm:2}
  $ \down{a,b}{a=f(a,b) \And b=g(a,b)} $
\Conn{ Lemma~\ref{BLK:lemma} with $F,G:=f,g$ }
  $ \down{a,b}{a=f(a,b) \And b=g(a,\bot)} $
\Conn{ substitute equals for equals }
  $ \down{a,b}{a=f(a,g(a,\bot)) \And b=g(a,\bot)} $
\Conn{ Lemma~\ref{BLK:lemma} with \nlhint
    $G,F:= \LambdaExpr{a,b}{}{f(a,g(a,\bot))},
           \LambdaExpr{a,b}{}{g(a,\bot)}$ }
  $ \down{a,b}{a=f(\bot,g(\bot,\bot)) \And b=g(a,\bot)} $
}
This calculation shows that an expression for the least solution of
$a$ in equation (\ref{spec:thm:2}) is
\[  f(\bot,g(\bot,\bot))  \]
By a symmetric argument, an expression for the least solution of $b$
in equation (\ref{spec:thm:2}) is
\[  g(f(\bot,\bot),\bot)  \]
That is, an expression for (\ref{spec:thm:2}) is
\[
  ( ~~
  f(\bot,g(\bot,\bot)), ~~
  g(f(\bot,\bot),\bot) ~~
  )
\]
which is the Pruned Closed Form.

Using the result for $\Bool^2$, we can show that the Pruned Closed
Form is also the least fixpoint in $\Bool^3$.  For any monotonic
boolean functions $f$, $g$, and $h$:
\Calc{
\calcEqLabel{spec:thm:3}
  $ \down{a,b,c}{a=f(a,b,c) \And b=g(a,b,c) \And c=h(a,b,c)} $
\Conn{ Lemma~\ref{BLK:lemma} with $G:=h$ (and with $F$ as the
  isomorphic \nlhint representation of functions $f$ and $g$) }
  $ \down{a,b,c}{a=f(a,b,c) \And b=g(a,b,c) \And c=h(a,b,\bot)} $
\Conn{ substitute equals for equals }
  $ \down{a,b,c}{a=f(a,b,c) \And b=g(a,b,h(a,b,\bot)) \And c=h(a,b,\bot)} $
\Conn{ Lemma~\ref{BLK:lemma} with
       $G:=\LambdaExpr{a,b,c}{}{g(a,b,h(a,b,\bot))}$ }
  $ \down{a,b,c}{a=f(a,b,c) \And b=g(a,\bot,h(a,\bot,\bot)) \And
                 c=h(a,b,\bot)} $
\Conn{ substitute equals for equals }
  $ \down{a,b,c}{\begin{array}[t]{l}
                 a=f(a,~ g(a,\bot,h(a,\bot,\bot)),~ c) \And \\
                 b=g(a,\bot,h(a,\bot,\bot)) \And
                 c=h(a,b,\bot)}
                 \end{array} $
\Conn{ the first 3 steps of this calculation, in reverse order }
  $ \down{a,b,c}{\begin{array}[t]{l}
                 a=f(a,~ g(a,\bot,h(a,\bot,\bot)),~ c) \And \\
                 b=g(a,b,c) \And
                 c=h(a,b,c)}
                 \end{array} $
\Conn{ Lemma~\ref{BLK:lemma} with $G:=g$ }
  $ \down{a,b,c}{\begin{array}[t]{l}
                 a=f(a,~ g(a,\bot,h(a,\bot,\bot)),~ c) \And \\
                 b=g(a,\bot,c) \And
                 c=h(a,b,c)}
                 \end{array} $
\Conn{ substitute equals for equals }
  $ \down{a,b,c}{\begin{array}[t]{l}
                 a=f(a,~ g(a,\bot,h(a,\bot,\bot)),~ c) \And \\
                 b=g(a,\bot,c) \And
                 c=h(a,g(a,\bot,c),c)}
                 \end{array} $
\Conn{ Lemma~\ref{BLK:lemma} with
       $G:=\LambdaExpr{a,b,c}{}{h(a,g(a,\bot,c),c)}$ }
  $ \down{a,b,c}{\begin{array}[t]{l}
                 a=f(a,~ g(a,\bot,h(a,\bot,\bot)),~ c) \And \\
                 b=g(a,\bot,c) \And
                 c=h(a,g(a,\bot,\bot),\bot)}
                 \end{array} $
\Conn{ substitute equals for equals }
  $ \down{a,b,c}{\begin{array}[t]{l}
                 a=f(a,~
                     g(a,\bot,h(a,\bot,\bot)),~
                     h(a,g(a,\bot,\bot),\bot)) \And
                 \\
                 b=g(a,\bot,c) \And
                 c=h(a,g(a,\bot,\bot),\bot)}
                 \end{array} $
\Conn{ Lemma~\ref{BLK:lemma} with
       $G:=$ \nlhint $\LambdaExpr{a,b,c}{}{
             f(a,~ g(a,\bot,h(a,\bot,\bot)),~ h(a,g(a,\bot,\bot),\bot))}$ }
  $ \down{a,b,c}{\begin{array}[t]{l}
                 a=f(\bot,~
                     g(\bot,\bot,h(\bot,\bot,\bot)),~
                     h(\bot,g(\bot,\bot,\bot),\bot)) \And \\
                 b=g(a,\bot,c) \And
                 c=h(a,g(a,\bot,\bot),\bot)}
                 \end{array} $
}
This calculation shows that an expression for the least solution of
$a$ in (\ref{spec:thm:3}) is
\[
  f(\bot,~
    g(\bot,\bot,h(\bot,\bot,\bot)),~
    h(\bot,g(\bot,\bot,\bot),\bot))
\]
and similarly for $b$ and $c$.

Our main result is that the Pruned Closed Form is the least fixpoint
in $\Bool^n$ for any $n$.  In the next section, we prove this result
directly, not using Lemma~\ref{BLK:lemma}.

\Section{The theorem}

We are given $n \atleast 1$ monotonic functions $f_1, \ldots, f_n \colon \Bool^n
\to \Bool$, where $\Bool$ is the boolean domain $\{0,1\}$ ordered by
$\atmost$ (with $0 \atmost 1$).  To represent an indexed $n$-tuple of
things, like a list of booleans $x_1,\ldots,x_n$, we write
$\tup{x}$.  The fact that the given functions are monotonic is written
as follows, for any index $i$ and any tuples of booleans $\tup{x}$ and
$\tup{y}$:
\[
  \tup{x} \tupatmost \tup{y}
  ~~\Imp~~
  f_i.\tup{x} \atmost f_i.\tup{y}
\]
where an infix dot (with the highest operator precedence) denotes
function application, and the order $\tupatmost$ is the component-wise
ordering of tuples:
\[
  \tup{x} \tupatmost \tup{y} 
  \Equiv
  \Forall{i}{}{
    x_i \atmost y_i
  }
\]

We are interested in viewing the functions as specifying a system of
equations, namely:
\begin{equation}\label{system}\begin{array}{rrcl}
x_1, \ldots, x_n :
&  x_1 & = & f_1.(x_1,~ \ldots,~ x_n) \\
&  \multicolumn{2}{l}{\ldots} \\
&  x_n & = & f_n.(x_1,~ \ldots,~ x_n)
\end{array}\end{equation}
where the variables to the left of the colon show the unknowns.
We take a tuple of functions $(f_1,~\ldots,~f_n)$, which we can also
write as $\tup{f}$, to itself be a
function, one which produces a tuple from the results of applying the
given argument to each of the functions.  For example, for the
functions given above and an argument $\tup{x}$, we have:
\[
  (f_1,~\ldots,~f_n).\tup{x}
  \Equal
  (f_1.\tup{x},~\ldots,~f_n.\tup{x})
\]
Thus, we can write the system (\ref{system}) of equations as:
\[\begin{array}{rl}
\tup{x} :
&  \tup{x} \Equal \tup{f}.\tup{x}
\end{array}\]
We are interested in the \emph{least} (in the sense of the ordering
$\tupatmost$) solution $\tup{x}$ that satisfies this equation.  That
is, we are interested in the \emph{least fixpoint} of the function
$\tup{f}$.  Because the lattice of boolean $n$-tuples has height $n$,
the least fixpoint of $\tup{f}$ can be reached by applying $\tup{f}$
$n$ times starting from the bottom element of the lattice.  That is,
the least fixpoint of $\tup{f}$ is given by:
\[  \tup{f}^n.\zerotup  \]
where exponentiation denotes successive function applications and
$\zerotup$ is the tuple of $n$ 0's.

%
% vkuncak: removed this old introduction
%
% If $n$ is large, say in the 100's or 1000's or even higher, we may not
% want to do all of these repeated applications of $\tup{f}$, especially
% if we expect each function $f_i$ to depend only on a small number of
% the elements in the argument tuple.  Instead, we can start from the
% tuple $\tup{x}$, repeatedly expanding occurrences of each $x_i$
% according to the corresponding equation in (\ref{system}), namely
% replacing $x_i$ with $f_i.\tup{x}$.  If, in this expansion, a term
% $x_i$ occurs as a subexpression of an argument to an application of
% function $f_i$, then this occurrence can be replaced by 0, rather than
% by $f_i.\tup{x}$.  Hence, a computation that performs the expansion
% eventually terminates.  We next prove that such a computation
% arrives at the correct result.

To precisely specify the Pruned Closed Form, we introduce a
notation that keeps track of which functions have been
applied in the enclosing context.  In particular, we use a
set that contains the indices of the functions already
applied.  Formally, we define the following family of
functions, for any index $i$ and set $S$ of indices:
\[
  g_{S,i} \Equal \left\{
  \begin{array}{ll}
    f_i \comp (g_{S \cup \{i\},1},~\ldots,~g_{S \cup \{i\},n})
        & \mbox{if $i \not\in S$} \\
    \LambdaExpr{\tup{x}}{}{0}
        & \mbox{if $i \in S$}
  \end{array}
  \right.
\]
Taking advantage of our previous notation and using $\zerofunc$ to
denote the function that always returns 0 (that is, the boolean 0
extended pointwise to a boolean function), we can write the definition
of $g$ as follows:
\[
  g_{S,i} \Equal \left\{
  \begin{array}{ll}
    f_i \comp \tup{g_{S \cup \{i\}}}
        & \mbox{if $i \not\in S$} \\
    \zerofunc
        & \mbox{if $i \in S$}
  \end{array}
  \right.
\]

Our goal is now to prove the following:
\begin{theorem}\label{main:theorem}
\[
  \tup{g_{\emptyset}}.\zerotup \Equal \tup{f}^n.\zerotup
\]
\end{theorem}

\Section{Proof}  \label{sec:proof}

We start by proving some lemmas that we use in the proof of this
theorem.

\begin{lemma}\label{lemma:1}
For any index $i$ and for any $S \subsetneqq \{1,\ldots,n\}$,
\[
  g_{S,i}.\zerotup ~~\atmost~~ (f_i \comp \tup{f}^{n-|S|-1}).\zerotup
\]
\end{lemma}
\Proof{}
By induction on $n-|S|$.  Let $T$ denote $S \cup \{i\}$.  We consider
three cases.
\Case{i \in S}
\Calc{
  $ g_{S,i}.\zerotup $
\Conn{ definition of $g$, since $i \in S$ }
  $ \zerofunc.\zerotup $
\Conn{ definition of $\zerofunc$ }
  $ 0 $
\conn{\atmost}{ $0$ is bottom element of $\atmost$ }
  $ (f_i \comp \tup{f}^{n-|S|-1}).\zerotup $
}
\Case{i \not\in S \And |T| < n}
\Calc{
  $ g_{S,i}.\zerotup $
\Conn{ definition of $g$, since $i \not\in S$ }
  $ (f_i \comp \tup{g_T}).\zerotup $
\Conn{}
  $ (f_i \comp (g_{T,1},~ \ldots,~ g_{T,n})).\zerotup $
\Conn{ distribute $.\zerotup$ }
  $ f_i.(g_{T,1}.\zerotup,~ \ldots,~ g_{T,n}.\zerotup) $
\conn{\atmost}{ for each index $j$, induction hypothesis with
        $i,S \Gets j,T$, since \nlhint $|S|+1 = |T| < n$;
        and monotonicity of $f_i$ }
  $ f_i.((f_1 \comp \tup{f}^{n-|S|-2}).\zerotup,~
         \ldots,~
         (f_n \comp \tup{f}^{n-|S|-2}).\zerotup) $
\Conn{ distribute $.\zerotup$ }
  $ (f_i \comp
        (f_1 \comp \tup{f}^{n-|S|-2},~
         \ldots,~
         f_n \comp \tup{f}^{n-|S|-2})).\zerotup $
\Conn{ distribute ${}\comp \tup{f}^{n-|S|-2}$ }
  $ (f_i \comp
        (f_1,~
         \ldots,~
         f_n) \comp \tup{f}^{n-|S|-2}).\zerotup $
\Conn{ exponentiation }
  $ (f_i \comp \tup{f}^{n-|S|-1}).\zerotup $
}
\Case{i \not\in S \And |T| = n}
\Calc{
  $ g_{S,i}.\zerotup $
\Conn{ see first 3 steps of previous case }
  $ f_i.(g_{T,1}.\zerotup,~ \ldots,~ g_{T,n}.\zerotup) $
\Conn{ for each index $j$, $j \in T$, so $g_{T,j} = \zerofunc$ }
  $ f_i.\zerotup $
\Conn{ $|S|=n-1$, so $f^{n-|S|-1}$ is the identity function }
  $ (f_i \comp f^{n-|S|-1}).\zerotup $
\EndProof
}

The following corollary of Lemma~\ref{lemma:1} proves one direction
of Theorem~\ref{main:theorem}.
\begin{corollary}\label{cor:1}
\[
  \tup{g_{\emptyset}}.\zerotup \tupatmost \tup{f}^n.\zerotup
\]
\end{corollary}
\Proof{}
\Calc{
  $ \tup{g_{\emptyset}}.\zerotup $
\Conn{}
  $ (g_{\emptyset,1}.\zerotup,~ \ldots,~ g_{\emptyset,n}.\zerotup) $
\conn{\tupatmost}{ for each index $j$, Lemma~\ref{lemma:1}
        with $i,S \Gets j,\emptyset$ }
  $ ((f_1 \comp \tup{f}^{n-1}).\zerotup,~
     \ldots,~
     (f_n \comp \tup{f}^{n-1}).\zerotup) $
\Conn{ distribute $.\zerotup$ and ${}\comp \tup{f}^{n-1}$ }
  $ ((f_1,~ \ldots,~ f_n) \comp \tup{f}^{n-1}).\zerotup $
\Conn{ exponentiation }
  $ \tup{f}^n.\zerotup $
\EndProof
}

To support the remaining lemmas, we define one more family of
functions.  For any index $i$ and set $S$ of indices,
\[
  h_{S,i} \Equal \left\{
  \begin{array}{ll}
    f_i
        & \mbox{if $i \not\in S$} \\
    \zerofunc
        & \mbox{if $i \in S$}
  \end{array}
  \right.
\]

\begin{lemma}\label{lemma:pre-2}
For any index $i$, monotonic function $H \colon \Bool^n \to \Bool^n$,
and $m \atleast 0$,
\[
  (f_i \comp H^m).\zerotup \Equal 0
  ~~\Imp~~
  \Forall{p}{0 \atmost p \atmost m}{(f_i \comp H^p).\zerotup \Equal 0}
\]
\end{lemma}
\Proof{}
We prove the term of the quantification as follows:
\Calc{
  $ (f_i \comp H^p).\zerotup $
\conn{\atmost}{ monotonicity of $f_i$ and $H$, since
                $\zerotup \tupatmost H^{m-p}.\zerotup$ }
  $ (f_i \comp H^p).(H^{m-p}.\zerotup) $
\Conn{}
  $ (f_i \comp H^m).\zerotup $
\Conn{ antecedent }
  $ 0 $
\EndProof
}

\begin{lemma}\label{lemma:2}
For any index $i$, set $S$ of indices, $m \atleast 0$, and $T = S \cup
\{i\}$,
\[
  (f_i \comp \tup{h_S}^m).\zerotup \Equal 0
  ~~\Imp~~
  \Forall{p}{0 \atmost p \atmost m}{
        \tup{h_S}^p.\zerotup \Equal \tup{h_T}^p.\zerotup}
\]
\end{lemma}
\Proof{}
If $i \in S$, then $S=T$ and the consequent follows
trivially.  For $i \not\in S$, we prove the term of the quantification
by induction on $p$.
\Case{p=0}
Trivial---exponentiation with 0 gives identity function.
\Case{p>0}
\Calc{
  $ \tup{h_S}^p.\zerotup $
\Conn{ exponentiation, since $p>0$ }
  $ \tup{h_S}.(\tup{h_S}^{p-1}.\zerotup) $
\Conn{ distribute $.(\tup{h_S}^{p-1}.\zerotup)$ }
  $ (h_{S,1}.(\tup{h_S}^{p-1}.\zerotup),~
     \ldots,~
     h_{S,n}.(\tup{h_S}^{p-1}.\zerotup)) $
\Conn{ for any index $j$,
        $h_{S,j}.(\tup{h_S}^{p-1}.\zerotup)
         =
         h_{T,j}.(\tup{h_S}^{p-1}.\zerotup)$, see below }
  $ (h_{T,1}.(\tup{h_S}^{p-1}.\zerotup),~
     \ldots,~
     h_{T,n}.(\tup{h_S}^{p-1}.\zerotup)) $
\Conn{ distribute $.(\tup{h_S}^{p-1}.\zerotup)$ }
  $ \tup{h_T}.(\tup{h_S}^{p-1}.\zerotup) $
\Conn{ induction hypothesis with $p \Gets p-1$ }
  $ \tup{h_T}.(\tup{h_T}^{p-1}.\zerotup) $
\Conn{ exponentiation }
  $ \tup{h_T}^p.\zerotup $
}
Now for the proof of the third step in the calculation above.
If $j \neq i$, then $j \in S \Equiv j \in T$, so $h_{S,j} = h_{T,j}$.
If $j=i$, then:
\Calc{
  $ h_{S,i}.(\tup{h_S}^{p-1}.\zerotup) $
\Conn{ definition of $h$, since $i \not\in S$ }
  $ f_i.(\tup{h_S}^{p-1}.\zerotup) $
\Conn{ Lemma~\ref{lemma:pre-2} with $H,p \Gets \tup{h_S},~p-1$,
      using the antecedent of \nlhint Lemma~\ref{lemma:2} to fulfill the
      antecedent of Lemma~\ref{lemma:pre-2} }
  $ 0 $
\Conn{ definition of $\zerofunc$ }
  $ \zerofunc.(\tup{h_S}^{p-1}.\zerotup) $
\Conn{ definition of $h$, since $i \in T$ }
  $ h_{T,i}.(\tup{h_S}^{p-1}.\zerotup) $
\EndProof
}

We need one more lemma.
\begin{lemma}\label{prop:3}
For any index $i$, set $S$ of indices, and $m$ satisfying $0 \atmost m
\atmost n-|S|$,
\begin{equation}\label{prop:3:formula}
  (h_{S,i} \comp \tup{h_S}^m).\zerotup
  ~~\atmost~~
  g_{S,i}.\zerotup
\end{equation}
\end{lemma}
\Proof{}
By induction on $m$.  We consider three cases.
\Case{i \in S}
\Calc{
  $ h_{S,i} \comp \tup{h_S}^m $
\Conn{ definition of $h$, since $i \in S$ }
  $ \zerofunc \comp \tup{h_S}^m $
\Conn{ $\zerofunc$ is left zero element of $\comp$ }
  $ \zerofunc $
\Conn{ definition of $g$, since $i \in S$ }
  $ g_{S,i} $
}
\Case{i \not\in S \And m=0}
\Calc{
  $ (h_{S,i} \comp \tup{h_S}^m).\zerotup $
\Conn{ exponentiation, since $m=0$ }
  $ h_{S,i}.\zerotup $
\Conn{ definition of $h$, since $i \not\in S$ }
  $ f_i.\zerotup $
\conn{\atmost}{ monotonicity of $f_i$, since
        $\zerotup \tupatmost \tup{g_T}.\zerotup$ }
  $ f_i.(\tup{g_T}.\zerotup) $
\Conn{ definition of $g$, since $i \not\in S$ }
  $ g_{S,i}.\zerotup $
}
\Case{i \not\in S \And m>0}
It suffices to prove that the left-hand side of (\ref{prop:3:formula})
is 0 whenever the right-hand side is 0.  Therefore, we assume the
latter to be 0:
\begin{equation}\label{s:0}
  g_{S,i}.\zerotup \Equal 0
\end{equation}
and prove the former to be 0:
\Calc{
  $ (h_{S,i} \comp \tup{h_S}^m).\zerotup $
\Conn{ definition of $h$, since $i \not\in S$ }
  $ (f_i \comp \tup{h_S}^m).\zerotup $
\Conn{ exponentiation, since $m>0$ }
  $ (f_i \comp
     (h_{S,1},~\ldots,~h_{S,n}) \comp
     \tup{h_S}^{m-1}).\zerotup $
\Conn{ distribute ${}\comp \tup{h_S}^{m-1}$ and $.\zerotup$ }
  $ f_i.((h_{S,1} \comp \tup{h_S}^{m-1}).\zerotup,~
         \ldots,~
         (h_{S,n} \comp \tup{h_S}^{m-1}).\zerotup) $
\conn{\atmost}{ (\ref{s:1}), see below; and monotonicity of $f_i$ }
  $ f_i.(g_{T,1}.\zerotup,~ \ldots,~ g_{T,n}.\zerotup) $
\Conn{}
  $ (f_i \comp \tup{g_T}).\zerotup $
\Conn{ definition of $g$, since $i \not\in S$ }
  $ g_{S,i}.\zerotup $
\Conn{ assumption (\ref{s:0}) }
  $ 0 $
}
In this calculation, we used the following fact:
for every index $j$,
\begin{equation}\label{s:1}
  (h_{S,j} \comp \tup{h_S}^{m-1}).\zerotup
  ~~\atmost~~
  g_{T,j}.\zerotup
\end{equation}
which we now prove.  We divide the proof of (\ref{s:1}) up into two
sub-cases.
\SubCase{(h_{S,j} \comp \tup{h_S}^{m-1}).\zerotup = 0}
Formula (\ref{s:1}) follows immediately.
\SubCase{(h_{S,j} \comp \tup{h_S}^{m-1}).\zerotup \neq 0}
First, we derive some consequences of assumption (\ref{s:0}):
\Calc{
  $ g_{S,i}.\zerotup \Equal 0 $
\conn{\imp}{ induction hypothesis with $S,i,m \Gets S,i,m-1$ }
\calcEqLabel{s:2}
  $ (h_{S,i} \comp \tup{h_S}^{m-1}).\zerotup \Equal 0 $
\Conn{ definition of $h$, since $i \not\in S$ }
  $ (f_i \comp \tup{h_S}^{m-1}).\zerotup \Equal 0 $
\conn{\imp}{ Lemma~\ref{lemma:2} with $m,p \Gets m-1,m-1$ }
\calcEqLabel{s:3}
  $ \tup{h_S}^{m-1}.\zerotup \Equal \tup{h_T}^{m-1}.\zerotup $
}
Now, calculating from the assumption we made in this sub-case:
\Calc{
  $ (h_{S,j} \comp \tup{h_S}^{m-1}).\zerotup \neq 0 $
\Conn{ (\ref{s:2}) }
  $ (h_{S,j} \comp \tup{h_S}^{m-1}).\zerotup \neq 0 \And i \neq j $

\conn{\imp}{ $i \neq j$, so $j \in S \Equiv j \in T$, so
    $h_{S,j} = h_{T,j}$ }
  $ (h_{T,j} \comp \tup{h_S}^{m-1}).\zerotup \neq 0 $
\Conn{ (\ref{s:3}) }
  $ (h_{T,j} \comp \tup{h_T}^{m-1}).\zerotup \neq 0$
\conn{\imp}{ induction hypothesis with $S,i,m \Gets T,j,m-1$ }
  $ g_{T,j}.\zerotup \neq 0 $
\conn{\imp}{}
  $ \mbox{(\ref{s:1})} $
}
This concludes the proof of Lemma~\ref{prop:3}.
\EndProof

And finally, the proof of the theorem:
\ProofOf{Theorem~\ref{main:theorem}}
The proof is a ping-pong argument.
\Calc{
  $ \tup{g_{\emptyset}}.\zerotup $
\conn{\tupatmost}{ Corollary~\ref{cor:1} }
\calcComment{\makebox[3cm]{---ping!}}
  $ \tup{f}^n.\zerotup $
\Conn{ exponentiation, since $n \atleast 1$ }
  $ ((f_1,~ \ldots,~ f_n) \comp \tup{f}^{n-1}).\zerotup $
\Conn{ distribute ${}\comp \tup{f}^{n-1}$ and $.\zerotup$ }
  $ ((f_1 \comp \tup{f}^{n-1}).\zerotup,~
     \ldots,~
     (f_n \comp \tup{f}^{n-1}).\zerotup) $
\Conn{ by definition of $h$, $h_{\emptyset,i} = f_i$ for each index
  $i$; and \nlhint thus also $\tup{h_{\emptyset}} = \tup{f}$ }
  $ ((h_{\emptyset,1} \comp \tup{h_{\emptyset}}^{n-1}).\zerotup,~
     \ldots,~
     (h_{\emptyset,n} \comp \tup{h_{\emptyset}}^{n-1}).\zerotup) $
\conn{\tupatmost}{ Lemma~\ref{prop:3} with $S,i,m \Gets
  \emptyset,i,n-1$ for each $i$ }
  $ (g_{\emptyset,1}.\zerotup,~
     \ldots,~
     g_{\emptyset,n}.\zerotup) $
\Conn{ distribute $.\zerotup$ }
\calcComment{\makebox[3cm]{---pong!}}
  $ \tup{g_{\emptyset}}.\zerotup $
\EndProof
}

\Section{Related Work and Acknowledgments}

Our theorem has already found a use, namely in the translation of
boolean programs into satisfiability formulas~\cite{Leino:SPIN2003}.

Before we knew of the \BL{} Theorem, one of us (Kuncak) proved the
theorem as detailed in Section~\ref{sec:proof}.  Tony Hoare then
proposed a way to prove the theorem in a way that would eliminate
recursive uses of variables, one by one.  In doing this, Hoare also
proved what essentially amounts to the \BL{} Theorem, appealing only
to the Tarski Fixpoint Theorem~\cite{Tarski:theorem}.  We elaborated
this format in Section~\ref{sec:bl}, to whose formulation Carroll
Morgan also contributed.  We learnt about the \BL{} Theorem from
Patrick Cousot.  The theorem is often called simply the Beki\'c
Theorem, but de~Bakker~\cite{deBakker:book} traces an independent
proof thereof to Leszczy{\l}owski.  Finally, we are grateful for
feedback from the Eindhoven Tuesday Afternoon Club and the participants
of the IFIP WG 2.3 meeting in Biarritz, France (March 2003).

\bibliography{krml118}

\end{document}